\documentclass[a4paper,onecolumn,oneside,fleqn,10pt]{article}
\usepackage{abstract}
\usepackage{amsmath}
\usepackage{authblk}
\usepackage[english]{babel}
\usepackage{bm}
\usepackage[font={color=blue},figurename=Figure,labelfont={it}]{caption}
\usepackage{cite}
\usepackage{enumitem}   
\usepackage{fchicago} 
\usepackage{float}
\usepackage[left=2cm,right=2cm,top=2cm,bottom=2cm]{geometry}
\usepackage{graphicx}
\usepackage{hyperref}
\usepackage{indentfirst}
\usepackage[utf8]{inputenc}
\usepackage[switch,columnwise]{lineno}
\usepackage{lipsum} 
\usepackage{pdfpages}
\usepackage{rotating}
\usepackage{setspace} 
\usepackage{subcaption}
\usepackage{newtxtext}
\usepackage[varg]{newtxmath}
\usepackage{wasysym}
\usepackage{fmtcount} 
\usepackage{chngcntr}
\usepackage{titlesec}
\usepackage{listings}
\usepackage[table]{xcolor}  
\usepackage{microtype}
\usepackage{footmisc} 

\counterwithin{equation}{section}
\renewcommand{\theequation}{%
  \thesection.%
  \ifnum\value{equation}<10 0\fi%
  \arabic{equation}%
}

\definecolor{blue}{RGB}{0,0,255}
\definecolor{matlabblue}{rgb}{0,0.4470,0.7410}

\hypersetup{
	colorlinks   = true,        
	linkcolor    = matlabblue,  
	citecolor    = matlabblue,  
	urlcolor     = matlabblue,  
	filecolor    = matlabblue,   
	linktoc      = all
}
\DeclareCaptionFont{bluefont}{\color{matlabblue}}
\newcommand{\via}{\textit{via }}

\newcommand{\eg}{\textit{eg. }}
\newcommand{\cf}{\textit{cf. }}
\newcommand{\ie}{\textit{ie. }}

\newcommand{\insitu}{\textit{in situ }}

\newcommand{\degree}{$^{\circ}$}

\usepackage{fancyhdr}
\graphicspath{{images/}{figures/pdf/}}

\counterwithout{equation}{section}       
\renewcommand{\theequation}{\arabic{equation}}

\begin{document}

\title{\huge On seasonal trunk thermal buffering and its electrical signature in mature trees}

\author[1]{Boulé Jean-Baptiste}
\author[2]{Courtillot Vincent}
\author[3]{Gibert Dominique}
\author[1]{Lopes Fernando}
\author[4]{Maineult Alexis}
\author[5]{Zuddas Pierpaolo}
\author[6]{de Bremond d'Ars Jean}

\affil[1]{\small Muséum National d’Histoire Naturelle, CNRS UMR7196, INSERM U1154, Paris, France}
\affil[2]{\small Académie des Sciences, Institut de France, Paris, France}
\affil[3]{\small DeepField Sensing, France}
\affil[4]{\small Laboratoire de Géologie de l’ENS, UMR 8538, Paris, France}
\affil[5]{\small Sorbonne Université, CNRS, METIS,UMR7619, Paris, France}
\affil[6]{\small Université de Rennes, CNRS, Géosciences Rennes, UMR 6118, Rennes, France}

\renewcommand\Authands{ and } 

\makeatletter
\renewcommand\AB@affilsepx{\\[0.5em]}            
\makeatother

\date{}
\maketitle

\newpage

\begin{abstract}
Tree trunks contain living tissues whose functioning depends on their thermal environment, yet the processes governing trunk temperature under field conditions remain poorly understood. In particular, it is unclear whether hydraulic transport contributes to buffering trunk temperature against atmospheric variability. We investigated this question by continuously monitoring sapwood temperature, local air temperature, soil temperature at 1~m depth, and spontaneous electrical potential (SP) in six mature trees, comprising three oaks and three hornbeams, over multiple years in a temperate urban forest garden. trunk temperature exhibited a smoother seasonal cycle than air temperature and remained consistently closer to deep-soil temperature, indicating substantial thermal buffering. Seasonal components extracted from the time series revealed a coherent delayed relationship between SP and the trunk-soil temperature difference. Phase-space analysis showed reproducible hysteresis across individuals, and instantaneous phase estimates indicated a lag of approximately 100~days in five of the six trees. A minimal energy-balance model further showed that trunk heat storage alone was insufficient to reproduce the observed seasonal dynamics. Positive effective soil-coupling coefficients were obtained for most individuals, although their magnitude varied markedly among trees. These results are consistent with a contribution of vertically mediated, hydraulically linked heat transfer to seasonal trunk thermal regulation. They further suggest that spontaneous electrical potentials may provide a non-invasive, integrative indicator of slow hydraulic and thermal processes within trees. Such thermo-hydraulic coupling may influence the thermal environment of the cambium and phloem and could therefore contribute to tree responses to seasonal heat and climatic variability.

	 
\end{abstract}

\newpage

\section{Introduction\label{sec:I}} 
Trees are long-lived rooted organisms exposed to substantial and often rapid fluctuations in environmental temperature. While the buffering role of the canopy through shading and transpiration has been extensively investigated, the thermal behavior of the trunk itself has received comparatively less attention. Yet the trunk constitutes a large, water-rich organ whose thermal inertia greatly exceeds that of leaves or small branches. Because living tissues such as the vascular cambium and phloem are embedded within this structure, their functioning depends partly on the thermal environment of the trunk. Temperature influences cambial reactivation and dormancy, cell division and enlargement, and the differentiation of newly formed xylem. Seasonal and short-term trunk temperature variations may therefore affect the timing and rate of wood formation, while freeze-thaw exposure can alter xylem hydraulic integrity and embolism risk. Understanding trunk thermal dynamics is consequently important for interpreting growth phenology, vascular functioning, and whole-tree responses to climatic stress (\eg \shortciteNP{sperry1992,anfodillo1998,grivcar2006,begum2013,steppe2015}).

Trunk temperature is shaped by multiple interacting processes. External forcing includes radiative exchange and convective heat transfer with ambient air, whereas internal processes involve the storage and redistribution of heat within wood tissues and sap. Importantly, xylem sap flow connects the trunk to deeper soil layers, where temperature remains comparatively stable throughout the year. Through this hydraulic connection, vertically transported water may act as a carrier of thermal energy, thereby coupling the trunk to a relatively stable underground thermal reservoir. Such a mechanism would imply that trunk temperature is not governed solely by passive surface exchange, but also by active hydraulic transport (\eg \shortciteNP{herzog1997,cermak2004,wang2023,roman2025}). By reducing the amplitude or rate of trunk temperature changes, such transport could stabilize the thermal environment experienced by cambial, phloem, and xylem tissues. This buffering may be particularly relevant during seasonal transitions and climatic extremes, when tissue temperature can constrain metabolic activity, wood formation, and hydraulic safety.

Despite its conceptual plausibility, direct field evidence for sustained thermo-hydraulic coupling in mature trees remains limited. Continuous \insitu measurements over seasonal timescales are rare, and separating passive thermal inertia from hydraulically mediated heat transport is challenging. If sap flow contributes to trunk thermal regulation, then changes in hydraulic activity should co-vary with trunk temperature dynamics. One potential non-invasive indicator of hydraulic activity is the spontaneous electrical potential (SP) generated by electrokinetic processes associated with sap movement within the xylem. This signal provides an integrated physical measure of water transport within the trunk (\eg \shortciteNP{doussan1998,gindl1999,gibert2006,lemouel2010,vandegehuchte2015,siddiq2018,lemouel2024}).

Here, we test the hypothesis that mature tree trunks function as vertically coupled thermo-hydraulic systems in which hydraulic transport contributes to internal heat redistribution. We combine multi-year measurements of sapwood, ambient-air, and deep-soil temperatures with continuous recordings of spontaneous electrical potentials around the trunk (\eg \shortciteNP{fensom1963,koppan2000}). Our objectives were to determine whether trunk temperature is buffered relative to air and coupled to deep-soil temperature, whether spontaneous electrical potential provides a non-invasive signature of the slow hydraulic dynamics associated with this coupling, and whether the observed thermal behaviour is compatible with an additional vertically mediated heat-transfer process. By linking trunk temperature, soil temperature, and bioelectric activity, we seek to identify a potentially overlooked component of tree thermal regulation, with possible consequences for the thermal environment of vascular tissues and tree resilience to climatic variability.

\section{Material and Methods \label{sec:II}}
The study was conducted in the Jardin écologique of the Jardin des Plantes, Museum National d’Histoire Naturelle, Paris, France (48.843\degree N, 2.356\degree E). The site is located in an urban environment but consists of semi-natural soil typical of temperate forest systems, with a maximum depth of approximately 2 m. Root systems are predominantly superficial and laterally developed. Six mature trees were monitored: three oaks (Quercus sp.; OAK01-03) and three hornbeams (Carpinus sp.; HB01-03). Trees were exposed to natural precipitation and standard horticultural maintenance practices. No artificial thermal manipulation was applied.

\begin{figure}[H]
	\centering
	\includegraphics[width=1\textwidth]{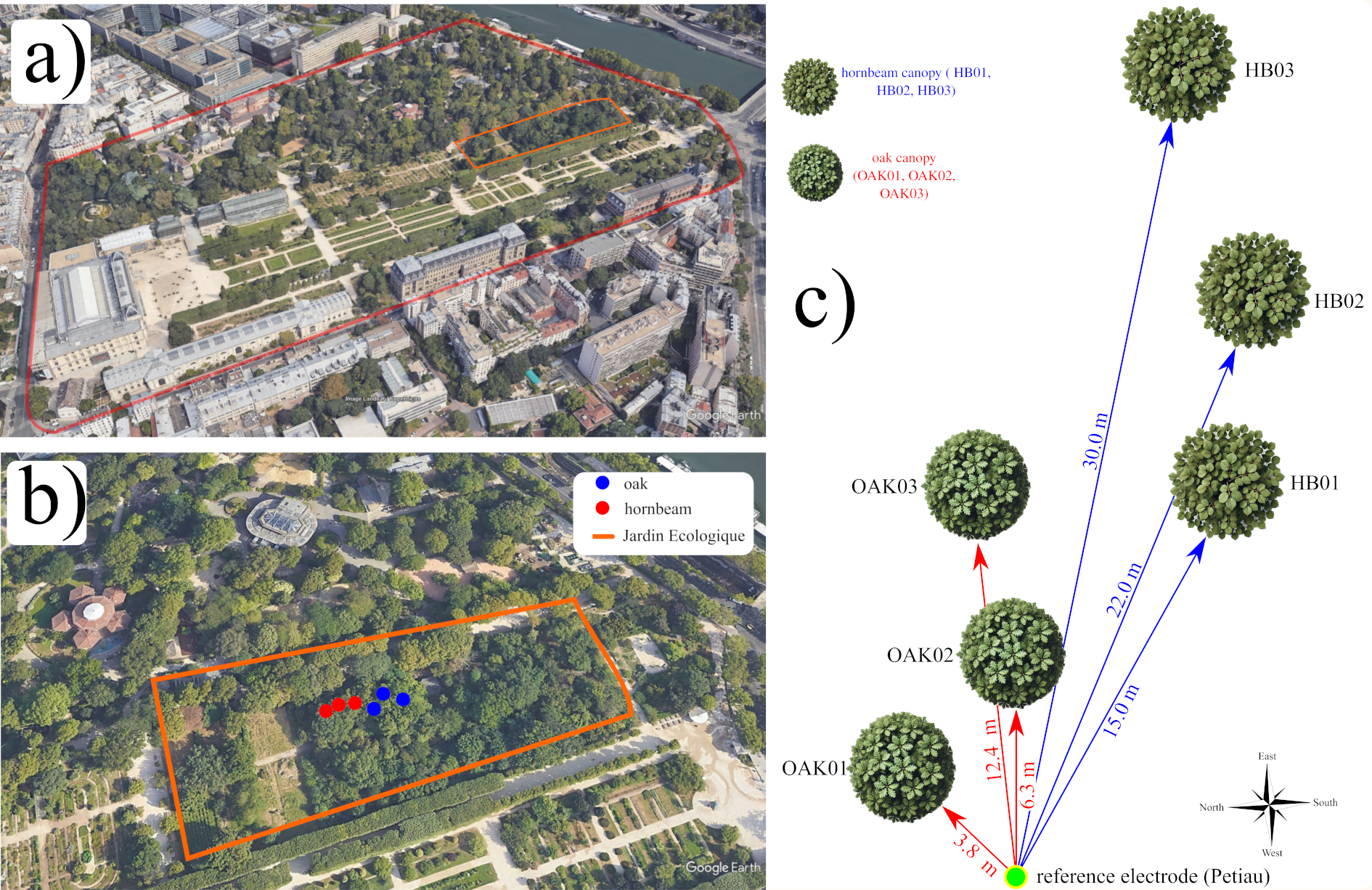}  
	\caption{Study site and experimental configuration. a) Aerial view of the Jardin des Plantes, Museum National d’Histoire Naturelle (MNHN, Paris, France) outlined in red. b) Zoom on the monitored area within the “Jardin écologique” (orange outline) showing the relative positions of the six instrumented trees: three oaks (blue) and three hornbeams (red). c) Schematic representation of tree positions relative to the non-polarizable KCl-filled Petiau reference electrode (SDEC; green), indicating tree-to-reference-electrode distances and orientation. North orientation is indicated. Imagery \copyright Google Earth}
	\label{Fig:01}
\end{figure}

The spatial configuration of the monitored trees and the position of the reference electrode are shown in Figure \ref{Fig:01}. The six instrumented trees were located within a restricted area of the Jardin écologique, with tree-to-reference-electrode distances ranging from 3.8 m to 30 m. The Petiau reference electrode (\eg \shortciteNP{petiau2000}) was installed at approximately 1 m depth in the soil near the first oak tree. The north-facing orientation of sensors was selected to minimize direct solar radiation effects and ensure consistent exposure across individuals.

Trunk temperature was measured using calibrated platinum resistance thermometers (Pt-100, $\pm$0.05\degree C accuracy). For each tree, two probes were inserted on the north-facing side of the trunk at heights of 50 cm and 100 cm above ground level. Probes were inserted approximately 1-2 cm into the sapwood. An additional Pt-100 sensor was positioned in the air near the trunk (at 75 cm above ground) to measure local ambient temperature. Soil temperature was recorded at 1 m depth using four Pt-1000 sensors distributed near the monitored trees. All sensors were laboratory-calibrated prior to deployment.

Spontaneous electrical potentials (SP) were recorded using stainless steel (316L) rod electrodes inserted 1-2 cm into the sapwood at 150 cm above ground level. Electrodes were oriented towards the four cardinal directions; analyses presented here focus on the north-facing electrodes co-located with thermal sensors. The non-polarizable Petiau reference electrode (SDEC; \shortciteNP{petiau2000}) was installed at approximately 1 m depth in the soil near the first oak tree.

Temperature and electrical-potential channels were digitized \insitu using Gantner A107 acquisition modules installed on each tree and connected to a multi-channel Gantner acquisition system. Digital data were transmitted to a central recording station through an RS485 bus. Measurements were recorded at 1 Hz with 24-bit resolution and high input impedance ($> 100 M\Omega$ ). Data were transmitted to a central recording station. Electrical potential data span from 19 December 2022 to 21 November 2025 for all six trees, with three short interruptions ($<10$ days each) between May and November 2023. Soil temperature measurements cover the same period. Trunk temperature recordings began on 3 February 2023 for OAK01 and on 17 January 2024 for the remaining trees.

Raw temperature and electrical-potential time series were down-sampled to daily means for seasonal analyses. Missing data segments corresponding to acquisition interruptions were excluded from analysis. Seasonal trends were extracted using Singular Spectrum Analysis (SSA; \eg \shortciteNP{vautard1992,lopes2024}). Phase relationships between spontaneous electrical potential and trunk-soil temperature differences ($\Delta T = T_{\mathrm{trunk}} - T_{\mathrm{soil}}$) were analysed using Lissajous representations and instantaneous phase estimation \via the Hilbert transform.

Because the daily records contained short-term meteorological fluctuations superimposed on a slower annual cycle, we used Singular Spectrum Analysis (SSA) to isolate the dominant seasonal component of each time series. SSA separates a record into a limited number of patterns derived directly from the data. It therefore allows the seasonal cycle to be reconstructed without assuming that it follows a sinusoidal shape or that its amplitude and timing are identical among trees. Only the reconstructed annual-scale components were used in the subsequent phase-space and Hilbert-transform analyses. Technical details of the decomposition and component selection are provided in Appendix~\ref{App:B}.

To visualize the seasonal relationship between the electrical and thermal signals, we used a phase-space hysteresis plot, also known as a Lissajous representation. Such plots provide a geometrical representation of the phase relationship between oscillatory variables; nearly synchronous signals produce a narrow trajectory, whereas a temporal offset opens the trajectory into a loop (\eg \shortciteNP{ewoldt2008}). Hysteresis plots have also been used in tree ecophysiology to identify delayed relationships between sap velocity and environmental or physiological drivers (\eg \shortciteNP{gimenez2019}). In this plot, the SSA-reconstructed spontaneous electrical potential was plotted against the corresponding SSA-reconstructed trunk-soil temperature difference. This representation provides an intuitive means of determining whether the two seasonal signals evolve simultaneously or with a delay. If both variables vary nearly synchronously, the trajectory collapses towards a narrow line. By contrast, if one variable responds later than the other, the trajectory forms an elliptical or loop-shaped path. The presence and width of such a loop therefore reveal a seasonal phase lag and hysteretic behaviour, consistent with a delayed response or memory within the coupled soil-tree system. The Lissajous representation was used as a qualitative visualization of this relationship, whereas the Hilberttransform analysis provided its quantitative expression as a time lag in days.

The duration of the seasonal offset was then quantified using instantaneous phase analysis. The Hilbert transform was applied separately to the SSA-reconstructed SP and trunk-soil temperature-difference signals to associate each seasonal cycle with an instantaneous phase. The phase difference between the two signals was converted into an equivalent time lag using a period of 365.25~days. With the convention adopted here,
\begin{equation*}
	\Delta \phi = \phi_{\mathrm{SP}}-\phi_{\Delta T},
\end{equation*}

negative lag values indicate that the seasonal evolution of SP occurred later than that of the trunk-soil temperature difference. The Hilbert analysis therefore provided a quantitative estimate, expressed in days, of the delay visualized qualitatively in the phase-space hysteresis plots.

To determine whether seasonal trunk-temperature buffering could be explained by heat storage alone, we formulated a minimal energy-balance model for a 1~m trunk segment. The model compares the rate at which heat is stored within the trunk with heat exchange with the surrounding air and with a possible effective contribution from the soil-root system. It is intentionally simplified and is used here as a diagnostic framework rather than as a complete description of radial and axial heat transport within the trunk. Conservation of energy gives,
\begin{equation}
	C\dfrac{dT}{dt}|_{trunk}=Q_{air} + Q_{soil},
\end{equation}

Here, $C$ is the effective heat capacity of the 1~m trunk segment, expressed in $\mathrm{J\,K^{-1}}$, and $Q_{\mathrm{air}}$ and $Q_{\mathrm{soil}}$ are net rates of heat transfer into the trunk, expressed in watts. With this sign convention, a positive value warms the trunk segment, whereas a negative value represents heat loss.

Heat exchange with the surrounding air was represented as,
\begin{equation}
	Q_{\mathrm{air}} = hA\left(T_{\mathrm{air}}-T_{\mathrm{trunk}}\right),
\end{equation}

where $h$ is an effective convective heat-transfer coefficient and $A$ is the lateral surface area of the trunk segment. This term is positive when the surrounding air is warmer than the trunk and negative when the trunk loses heat to the atmosphere.

A second term was introduced to represent the possible contribution of the soil-root system,
\begin{equation}
	Q_{\mathrm{soil}} = K_s\left(T_{\mathrm{soil}}-T_{\mathrm{trunk}}\right),
\end{equation}

where $T_{\mathrm{soil}}$ is approximated by the temperature measured at 1~m depth. The effective coefficient $K_s$, expressed in $\mathrm{W\,K^{-1}}$, summarizes the strength of the thermal connection between the trunk and the belowground reservoir. It should not be interpreted as a purely conductive coefficient, because it may integrate conductive, hydraulically advective, and root-soil contributions. 

The model was fitted independently for each tree. Its purpose was not to derive a direct measure of sap-mediated heat transport, but to test whether trunk heat storage and atmospheric exchange alone were sufficient to reproduce the observed seasonal trunk-temperature dynamics.

\section{Results \label{sec:III}}
\begin{figure}[H]
	\centering
	\includegraphics[width=1\textwidth]{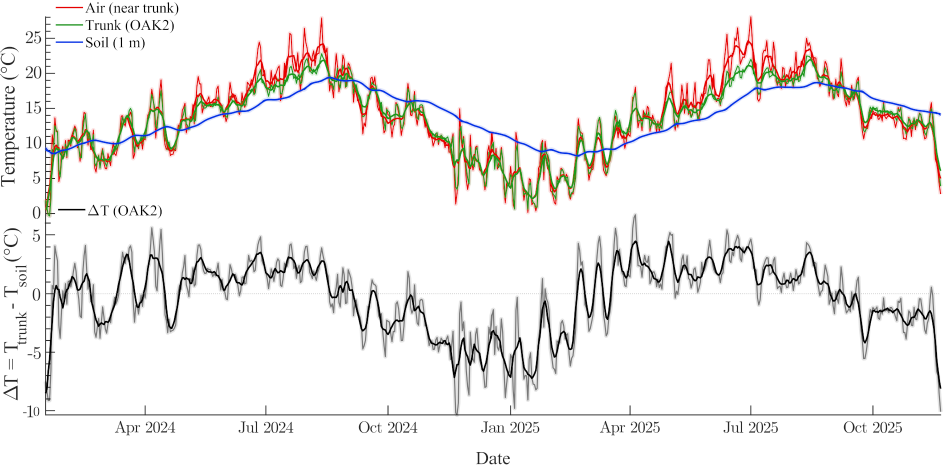}  
	\caption{Seasonal thermal dynamics of air, trunk and soil temperatures in OAK02.
		Upper panel: daily mean air temperature measured near the trunk, trunk temperature at 100 cm height, and soil temperature at 1 m depth from January 2024 to November 2025. Lower panel: air-trunk temperature difference, $\Delta T_{air-trunk}=T_{air}-T_{trunk}$. Thin lines represent daily means; thicker lines indicate a 7-day moving average. Short-term and seasonal variations in air temperature are attenuated in the trunk, illustrating the thermal buffering experienced by internal trunk tissues.}
	\label{Fig:02}
\end{figure}

At the seasonal scale, trunk temperature varied more smoothly and over a narrower range than local air temperature, while remaining closer to the temperature measured at 1~m soil depth (\cf Figure~\ref{Fig:02}). Short-term atmospheric fluctuations and seasonal air-temperature extremes were therefore attenuated within the sapwood.

Across the 2024 and 2025 annual cycles, the trunk-soil temperature difference ($\Delta T=T_{\mathrm{trunk}}-T_{\mathrm{soil}}$) changed gradually, with positive values during the warmer part of the year and negative values during colder periods. By contrast, rapid variations in air temperature were only weakly expressed in the trunk-temperature record. The recurrence of this pattern over two consecutive years demonstrates persistent seasonal thermal buffering of the trunk. This observation alone does not identify the underlying heat-transfer mechanism, which was examined subsequently using the electrical, phase, and energy-balance analyses.

The repetition of these patterns across consecutive years supports the robustness of the observed thermal coupling and suggests that trunk temperature dynamics are not solely governed by surface exchange with ambient air.

Full multi-year recordings for all six monitored trees are provided in  Figure \ref{Fig:S01} (\cf Appendix \ref{App:A}), illustrating the continuity and consistency of the thermal and electrical signals across individuals.

\begin{figure}[H]
	\centering
	\includegraphics[width=1\textwidth]{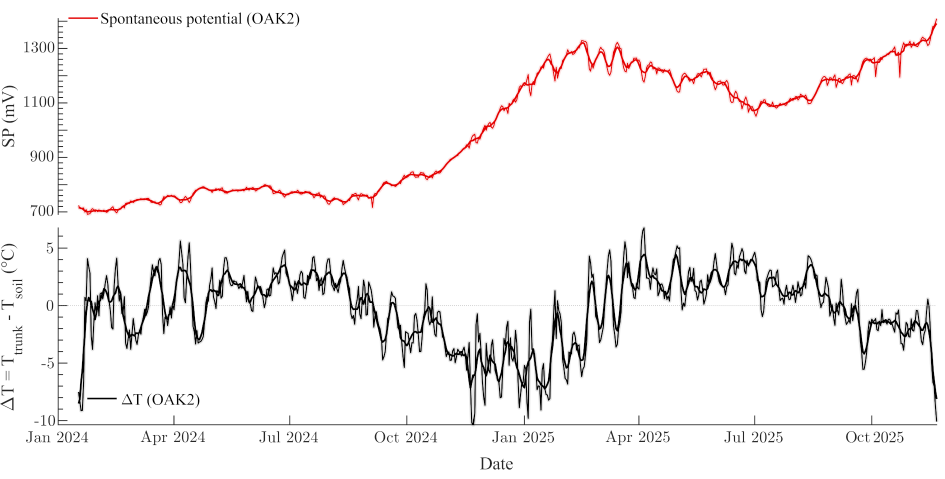}  
	\caption{Seasonal co-variation between spontaneous electrical potential and trunk thermal state in OAK02. Daily mean spontaneous electrical potential (SP, red) recorded at 150 cm height relative to the soil reference electrode and trunk-soil temperature difference ($\Delta T$, black) from January 2024 to November 2025. Thin lines represent daily means; thicker lines indicate a 7-day moving average to guide the eye. Both variables exhibit recurrent seasonal variability across two consecutive annual cycles, although their extrema do not occur synchronously.}
	\label{Fig:03}
\end{figure}

Having established that trunk temperature was seasonally buffered relative to air temperature, we next examined whether this thermal behaviour was accompanied by a coherent electrical signature. Spontaneous electrical potential exhibited pronounced seasonal variability over the 2024-2025 monitoring period (\cf Figure~\ref{Fig:03}). Its temporal evolution showed a recurrent seasonal relationship with the trunk-soil temperature difference ($\Delta T$) across both annual cycles. Periods of sustained positive or negative $\Delta T$ were accompanied by gradual changes in SP, although the two variables did not reach their seasonal extrema simultaneously. The recurrence of this relationship over consecutive years indicates that the electrical variability was associated with a persistent seasonal process rather than with isolated events. Short-term fluctuations in $\Delta T$ were not systematically mirrored by SP, suggesting that the electrical signal reflects processes operating over longer timescales than daily atmospheric variability.

Seasonal components reconstructed using Singular Spectrum Analysis (SSA) confirmed the presence of recurrent annual-scale variability across individuals and species (\cf Figure~\ref{Fig:S02}, Appendix~\ref{App:B}). We then examined whether the reconstructed electrical and thermal signals evolved synchronously or exhibited a systematic seasonal delay.

\begin{figure}[H]
	\centering
	\includegraphics[width=1\textwidth]{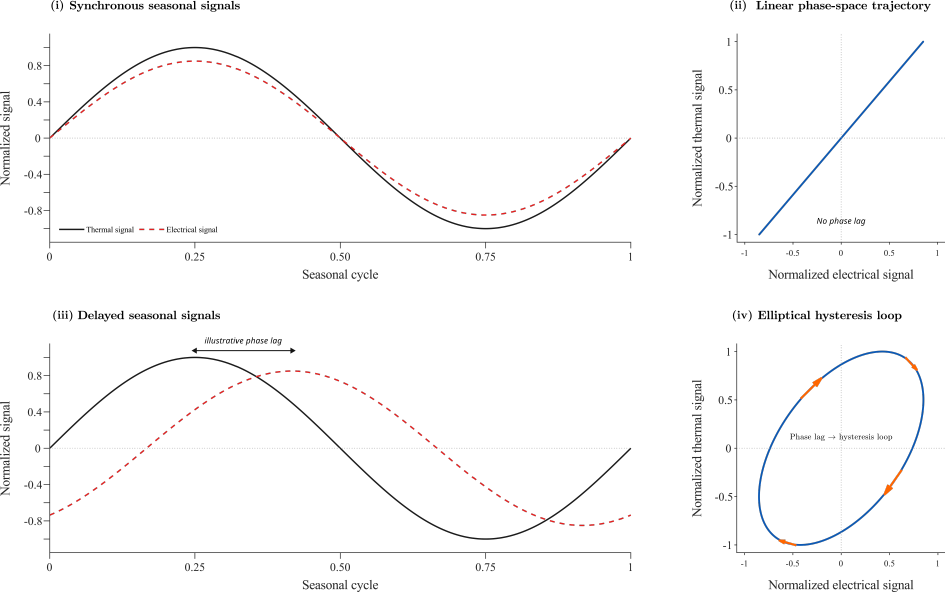}  
	\caption{
	Conceptual interpretation of a seasonal phase-space hysteresis plot. \textbf{(i)} Idealized thermal and electrical signals varying synchronously over one seasonal cycle. Their maxima, minima, and zero crossings occur simultaneously.
	\textbf{(ii)} When the normalized thermal signal is plotted against the normalized electrical signal, this synchronous behaviour produces a narrow linear trajectory, indicating the absence of a phase lag. \textbf{(iii)} Idealized seasonal signals in which the electrical response is delayed relative to the thermal signal.\textbf{(iv)} The corresponding phase-space trajectory forms an elliptical hysteresis loop. Arrows indicate the direction of progression through the seasonal cycle. The formation of a loop provides a qualitative signature of a phase shift and delayed system response; the lag itself is subsequently quantified using the Hilbert transform. All signals shown here are normalized and illustrative rather than experimental.
	}
	\label{Fig:04a}
\end{figure}

Before examining the measured relationship between SP and trunk-soil temperature difference, the interpretation of a phase-space hysteresis plot is illustrated conceptually in Figure~\ref{Fig:04a}. When the thermal and electrical seasonal signals vary synchronously, their phase-space trajectory collapses towards a narrow line (Figure~\ref{Fig:04a}i-ii). By contrast, a systematic delay of one signal relative to the other opens the trajectory into an elliptical loop (Figure~\ref{Fig:04a}iii-iv). Such a loop therefore provides an intuitive visual signature of seasonal phase lag and hysteretic behaviour, whereas its precise temporal magnitude cannot be inferred from the loop geometry alone and was quantified independently using instantaneous phase analysis.

\begin{figure}[H]
	\centering
	\includegraphics[width=1\textwidth]{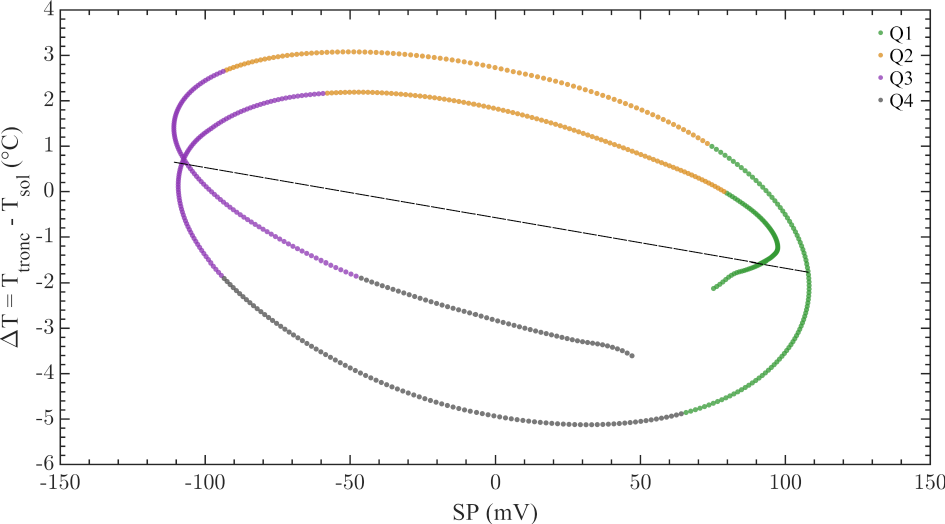}  
	\caption{ Seasonal phase-space hysteresis plot between spontaneous electrical potential (SP) and trunk-soil temperature difference ($\Delta T=T_{\mathrm{trunk}}-T_{\mathrm{soil}}$) for OAK02. The seasonal components of both signals were reconstructed using Singular Spectrum Analysis (SSA; \cf Figure~\ref{Fig:S02}, Appendix~B). Points are coloured according to calendar quarters (Q1-Q4). The closed trajectory indicates that the electrical and thermal seasonal signals do not evolve synchronously. Its loop-shaped geometry provides a qualitative representation of their phase relationship but does not directly determine the duration of the lag, which was estimated independently using instantaneous phase analysis.
}
	\label{Fig:04}
\end{figure}

The corresponding phase-space hysteresis plot derived from the measured signals of OAK02 is shown in Figure~\ref{Fig:04}. This representation shows how the electrical state of the trunk evolves relative to its thermal state over the seasonal cycle. The trajectory forms a closed loop rather than the narrow linear pattern expected for synchronous signals. Colouring the observations by calendar quarter reveals an organized annual progression, with distinct branches associated with the warming and cooling portions of the cycle. The loop therefore indicates that SP and trunk-soil temperature difference did not evolve synchronously. Its geometry provides a qualitative representation of their delayed seasonal relationship, but does not by itself determine the duration of the lag.

Because phase-space interpretation requires both variables to share a dominant seasonal timescale, the analysis was applied to the SSA-reconstructed annual components (\cf Figure~\ref{Fig:S02}, Appendix~\ref{App:B}). The raw SP and temperature records also contain daily variability and short-term meteorological fluctuations that can obscure the slower seasonal relationship. SSA reconstruction reduces the influence of these higher-frequency variations while preserving the timing and shape of the dominant annual-scale variability. The phase-space plot should therefore be interpreted as a representation of the relationship between the reconstructed seasonal components rather than between the unprocessed signals.

Similar closed or distorted phase-space loops were observed across the monitored trees, although their amplitude and geometry varied among individuals (\cf Figure~\ref{Fig:S03}, Appendix~\ref{App:C}). This recurrence indicates that a delayed seasonal relationship between SP and trunk thermal state was not restricted to OAK02, while the observed inter-individual differences suggest variability in the underlying physiological or environmental controls.

\begin{figure}[H]
	\centering
	\includegraphics[width=1\textwidth]{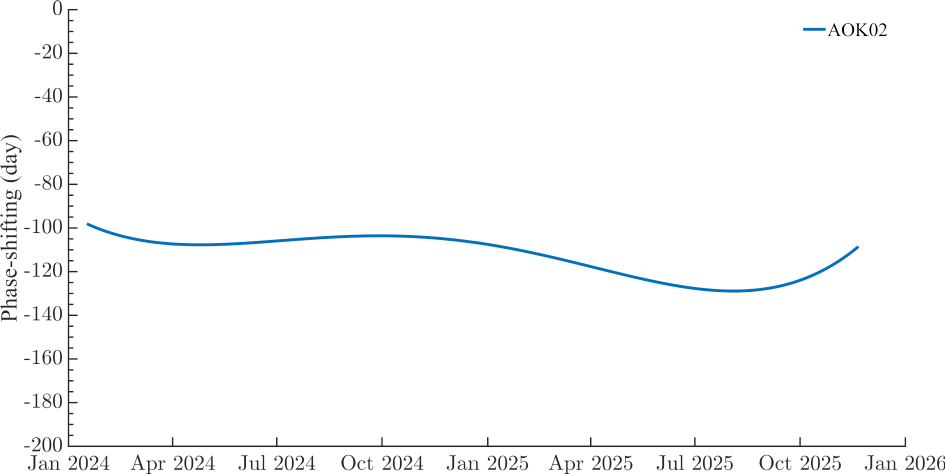}  
	\caption{Instantaneous seasonal phase lag between spontaneous electrical potential (SP) and trunk–soil temperature difference ($\Delta T$) for OAK02, estimated using the Hilbert transform. The phase lag remains consistently close to -100 days over the annual cycle, with only moderate intra-seasonal modulation. This persistent delay indicates that the electrical and thermal signals integrate slowly varying seasonal processes rather than responding synchronously. The lag is consistent with vertical thermo-hydraulic coupling, but should not be interpreted as a direct travel time for water or heat. Comparable phase lags are observed in all trees except OAK01 (\cf Figure \ref{Fig:S04}, Appendix \ref{App:D}).}
	\label{Fig:05}
\end{figure}

Because the geometry of the hysteresis loop indicates the presence of a delay but does not provide its duration directly, we then quantified the phase lag using instantaneous phase analysis. Instantaneous phase analysis using the Hilbert transform reveals a consistent seasonal lag of approximately -100 days between SP and trunk-soil temperature difference in OAK02 (\cf Figure \ref{Fig:05}). The lag remains remarkably stable over the annual cycle, with only moderate intra-seasonal modulation. Similar delays are observed in all trees except OAK01 (\cf Figure \ref{Fig:S04}, Appendix \ref{App:D}). Such a persistent phase shift corresponds to roughly one quarter of the annual cycle and indicates a slow dynamical response. This finding quantitatively confirms the hysteretic behavior observed in phase space (\cf Figure \ref{Fig:04}) and supports the interpretation of vertically mediated thermo-hydraulic coupling within the trunk.

Because the geometry of the hysteresis loop indicates the presence of a delay but does not provide its duration directly, we quantified the seasonal offset using instantaneous phase analysis. For OAK02, the phase lag remained close to $-100$~days over most of the annual cycle (\cf Figure~\ref{Fig:05}), with only moderate intra-seasonal variability. Under the sign convention adopted here, negative values indicate that the seasonal evolution of SP occurred later than that of the trunk-soil temperature difference. Broadly similar offsets were observed in five of the six monitored trees (\cf Figure~\ref{Fig:S04}, Appendix~\ref{App:D}). This recurrent delay confirms that the electrical and thermal seasonal signals did not evolve synchronously, but it should not be interpreted as a direct travel time for water or heat.

\begin{figure}[H]
	\centering
	\includegraphics[width=1\textwidth]{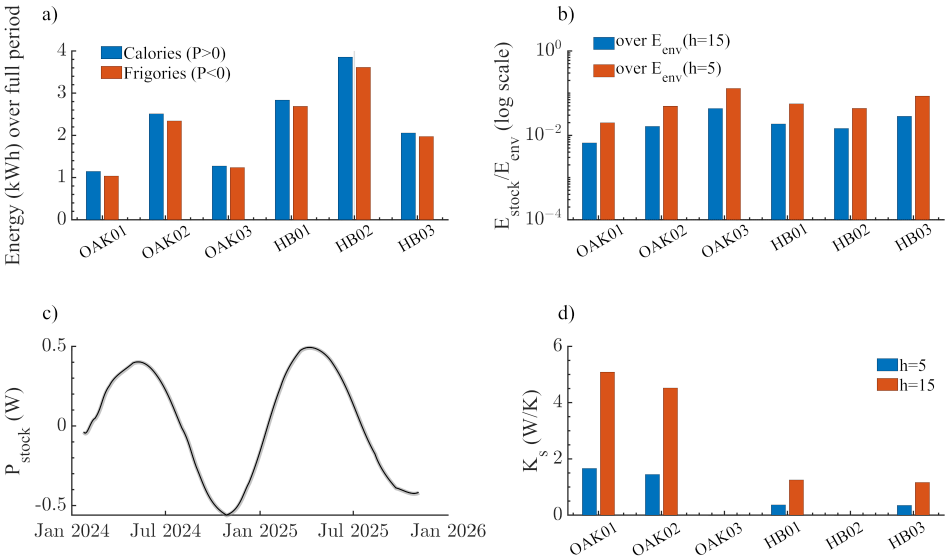}  
	\caption{Energetic constraints from a minimal trunk energy balance model (SSA seasonal signals). a) Integrated seasonal heat storage within a 1 m trunk segment, computed from measured temperature variations and estimated thermal capacity. Positive and negative contributions represent seasonal heat accumulation and heat release by the trunk, respectively. b) Ratio between integrated trunk heat storage and the equivalent convective exchange energy with ambient air for convective coefficients $h\in [5,15] \,  W\cdot m^{-2} \cdot K^{-1}$ (logarithmic scale). Ratios well below unity indicate that thermal inertia alone cannot explain seasonal buffering. c) Instantaneous trunk heat storage power $P_{stock}=C \dfrac{dT}{dt}$ for a representative tree (\ie OAK02), showing alternating seasonal heat accumulation and release. d) Effective soil coupling coefficient Ks obtained by fitting a minimal linear energy balance $C\dfrac{dT}{dt}=hA(T_{air} -T_{trunk})+ K_{s}(T_{soil}-T_{trunk})$. Positive fitted values are consistent with an additional effective thermal connection between the trunk and the belowground reservoir.}
	\label{Fig:06}
\end{figure}

Finally, we tested whether the observed seasonal buffering could be explained by trunk heat storage alone or whether an additional heat-transfer pathway was required. Seasonal trunk heat storage was estimated from the SSA-reconstructed temperature component. The integrated seasonal heat exchange within a 1 m trunk segment reaches values between approximately 1 and 4 kWh depending on tree size (\cf Figure \ref{Fig:06}a). The near balance between positive and negative contributions indicates a quasi-stationary seasonal regime. However, when compared to the equivalent convective energy exchange with ambient air, trunk heat storage represents only a small fraction of the atmospheric forcing (\cf Figure \ref{Fig:06}b). The ratio $E_{stock}/E_{env}$ remains well below unity ($10^{-4} - 10^{-1}$ range), demonstrating that thermal inertia alone cannot account for the observed seasonal buffering. The instantaneous storage power (\cf Figure \ref{Fig:06}c) reveals alternating periods of heat accumulation and release, consistent with the phase lag identified in the Lissajous analysis. To evaluate the role of vertical heat transfer, we fitted a minimal energy balance model including atmospheric exchange and a soil coupling term. The fitted values of the effective coupling coefficient $K_s$ were positive for most individuals, although their magnitude varied markedly among trees (\cf Figure~\ref{Fig:06}d). This pattern is consistent with an additional vertically mediated thermal contribution, while highlighting substantial inter-individual variability. Together, these results show that seasonal trunk temperature dynamics cannot be interpreted as a purely passive atmospheric response but reflect coupled soil-atmosphere energy transfer processes.

\section{Discussion \label{sec:IV}}
The present study provides multi-year field evidence that seasonal trunk-temperature dynamics are consistent with vertical thermo-hydraulic coupling in mature trees. Across six individuals and two species, trunk temperature exhibited reduced seasonal amplitude relative to air temperature and remained consistently closer to soil temperature (\cf Figure \ref{Fig:02}). Spontaneous electrical potentials co-varied coherently with trunk-soil temperature differences (\cf Figure \ref{Fig:03}), forming structured seasonal hysteresis loops in phase space (\cf Figure \ref{Fig:04} and Figure \ref{Fig:S03}). Energetic analysis further demonstrated that trunk thermal inertia alone cannot account for the observed seasonal buffering (\cf Figure~\ref{Fig:06}b). Positive fitted values of the effective soil coupling coefficient $K_s$ were obtained for most individuals, with marked inter-individual variation in magnitude (\cf Figure~\ref{Fig:06}d).

The phase-space analysis revealed a stable seasonal lag between electrical potential and thermal state, indicating that the two signals did not evolve synchronously. An instantaneous response to atmospheric temperature alone would not be expected to produce such a persistent loop. Instead, the observed hysteresis suggests that trunk electrical and thermal dynamics integrate processes operating over longer timescales. One plausible contribution is heat transport associated with xylem water movement, which connects the trunk to belowground water sources whose temperature varies more slowly than that of the atmosphere. Because water has a high heat capacity, hydraulic transport may contribute to the redistribution of heat within the soil-root-stem continuum. In this interpretation, the fitted coefficient $K_s$ represents an integrated thermal connection rather than a direct measurement of sap-mediated heat flux.

Instantaneous phase analysis quantified the seasonal offset between SP and trunk thermal state at approximately 100~days in five of the six monitored trees. The persistence of this delay indicates that the two signals integrate slowly varying seasonal processes rather than responding directly to short-term atmospheric fluctuations. This lag should not, however, be interpreted as the travel time of water or heat from the soil to the trunk. Instead, it characterizes the phase relationship between two seasonal signals that may both be influenced by hydraulic activity, soil temperature, phenology, and other environmental drivers. Its consistency across individuals supports the presence of a recurrent system-level process, while not identifying a unique underlying mechanism.

Previous studies have documented the influence of temperature on cambial activity and xylem processes, and sap flow measurements have highlighted the role of water transport in trunk heat dynamics. However, direct evidence of sustained seasonal thermo-hydraulic coupling in mature trees under natural conditions has remained limited.

Our results extend previous sap-flow-based approaches by incorporating continuous measurements of spontaneous electrical potential as a non-invasive indicator of whole-stem dynamics. The coherent seasonal relationship between SP and trunk thermal state suggests that electrical signals integrate slow processes associated with water transport and tree phenology. SP should not be interpreted as a direct measure of sap velocity, but it may provide a useful complementary signal for monitoring seasonal changes in tree hydraulic functioning.

Thermo-hydraulic coupling may contribute to stabilizing the thermal environment experienced by the vascular cambium, phloem, and differentiating xylem. Because cambial reactivation, cell division, cell enlargement, and xylem differentiation are temperature sensitive, even moderate buffering of trunk temperature could influence the timing and rate of wood formation. A reduced exposure to rapid atmospheric temperature changes may also limit thermal stress on living trunk tissues and help preserve xylem hydraulic integrity during seasonal transitions. Inter-individual differences in the fitted coupling coefficient (\cf Figure~\ref{Fig:06}d) may reflect variation in sapwood properties, root architecture, hydraulic activity, or local soil conditions. Such differences could contribute to contrasting capacities among trees to maintain trunk thermal stability during heat, cold, or drought events.

Several limitations constrain the mechanistic interpretation of these results. The minimal energy-balance model does not resolve radial or axial temperature gradients within the trunk, and the effective coefficient $K_s$ combines potentially conductive, advective, and structural contributions. In addition, SP is an integrative electrical signal rather than a direct measurement of sap velocity, and both SP and trunk temperature may respond to common seasonal drivers. Soil temperature measured at 1~m depth may also represent only part of the thermal environment explored by the root system. The observed relationships should therefore be interpreted as evidence consistent with thermo-hydraulic coupling rather than as a direct measurement of sap-mediated heat transport. Future studies combining direct sap-flow measurements, xylem water potential, soil moisture, and radial trunk-temperature profiles would allow the hydraulic and thermal contributions to be separated more explicitly.

The combination of continuous thermal monitoring and spontaneous electrical potential measurements provides a novel framework for studying whole-tree energy transport. Extending this approach across climatic gradients and species with contrasting hydraulic strategies may clarify the role of thermo-hydraulic coupling in tree resilience. Beyond plant physiology, the results highlight how bioelectric signals can serve as integrative indicators of ecosystem-scale physical processes.
	
\section{Conclusion \label{sec:V}}	
This study provides multi-year field evidence that seasonal trunk-temperature dynamics are consistent with vertical thermo-hydraulic coupling in mature trees.

Phase-space analysis demonstrated a stable seasonal hysteresis between spontaneous electrical potential and trunk-soil temperature difference, indicating delayed system response and memory effects. Energetic constraints further showed that trunk thermal inertia alone is insufficient to explain the observed buffering. A minimal energy balance model yielded positive effective soil coupling coefficients for most individuals, although their magnitude varied markedly among trees. These results are consistent with an additional vertically mediated contribution to trunk heat transfer.

Together, these observations support the hypothesis that hydraulic transport contributes to seasonal heat redistribution within tree trunks, in addition to its established role in water and nutrient transport. Such coupling could influence the thermal environment experienced by the vascular cambium, phloem, and differentiating xylem, and may therefore contribute to tree responses to seasonal and climatic variability. Continuous SP measurements may provide a useful complementary indicator of these slow whole-stem dynamics, although direct sap-flow and temperature-profile measurements will be required to resolve the underlying mechanisms.

By revealing a persistent seasonal phase delay and energetic constraints that cannot be explained by thermal inertia alone, this study provides evidence consistent with hydraulically mediated heat redistribution, challenging the interpretation of trunk temperature as a purely atmospheric signal.
\newpage
\bibliographystyle{fchicago}
\bibliography{thermo_arbres.bib}
\newpage

\counterwithin{equation}{section}        
\renewcommand{\theequation}{%
	\thesection.\ifnum\value{equation}<10 0\fi\arabic{equation}%
}

\appendix
\titleformat{\section}
{\normalfont\Large\bfseries}       
{Appendix~\thesection}             
{1em}                              
{}                 

\setcounter{figure}{0}
\renewcommand{\thefigure}{A\padzeroes[2]{\arabic{figure}}}
\section{Multi-year monitoring data\label{App:A}}
\begin{figure}[H]
	\centering
	\includegraphics[width=1\textwidth]{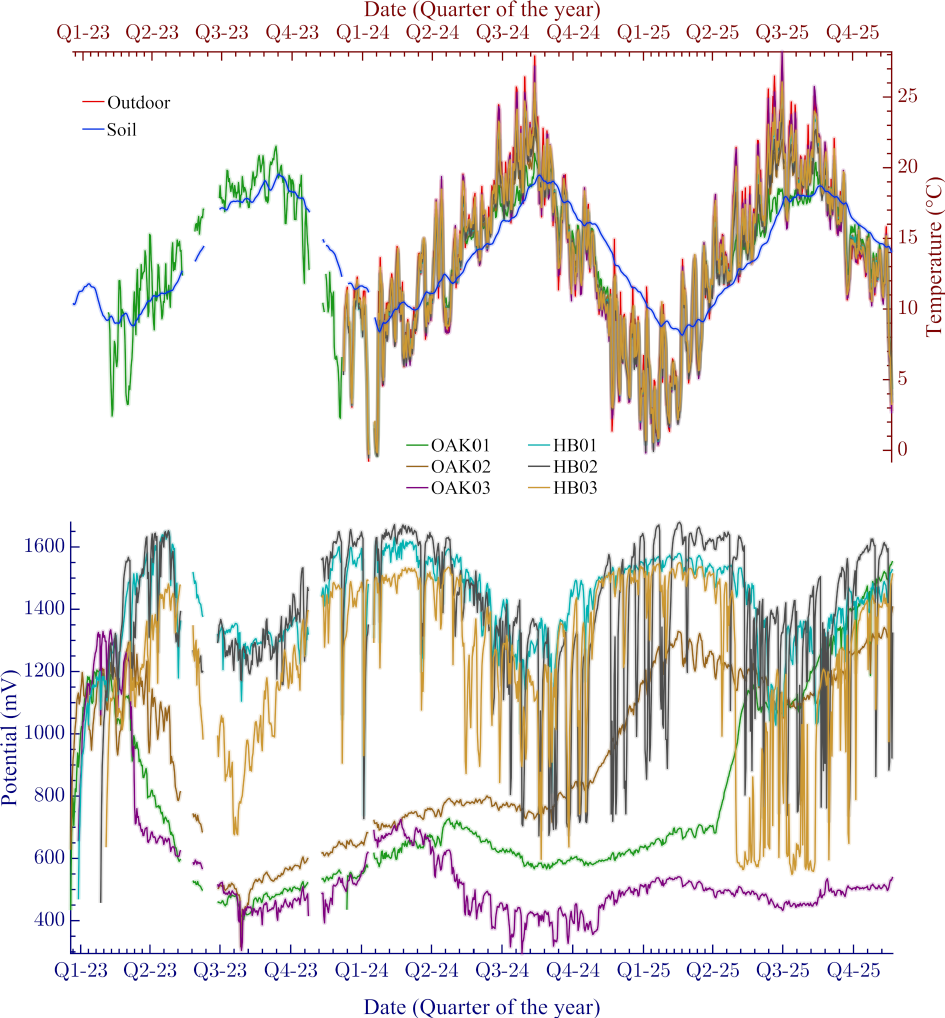}  
	\caption{Multi-year thermal and electrical monitoring of all six instrumented trees. Upper panel: daily mean air temperature (red), soil temperature at 1 m depth (blue), and trunk temperatures for three oaks (OAK01-03) and three hornbeams (HB01-03) from December 2022 to November 2025. Lower panel: corresponding spontaneous electrical potentials (SP, in mV) recorded at 150 cm height relative to a soil-installed Petiau reference electrode. Short data gaps correspond to brief acquisition interruptions ($<10$ days) between May and November 2023. Measurements continued beyond November 2025 but were truncated for analysis consistency.}
	\label{Fig:S01}
\end{figure}

Figure \ref{Fig:S01} presents the complete daily mean temperature and spontaneous electrical potential recordings for all six monitored trees over the 2022-2025 observation period. The upper panel shows ambient air temperature, soil temperature at 1 m depth, and trunk temperatures for three oaks and three hornbeams. The lower panel displays the corresponding spontaneous electrical potentials measured relative to a non-polarizable Petiau reference electrode.
	
The long-term continuity of the recordings highlights the robustness of the observed seasonal patterns across individuals and species. Minor data gaps correspond to short acquisition interruptions ($<10$ days) during 2023. These interruptions were excluded from subsequent analyses. The full multi-year dataset demonstrates the stability and reproducibility of both thermal and electrical signals over consecutive annual cycles.
\newpage 
\setcounter{figure}{0}
\renewcommand{\thefigure}{B\padzeroes[2]{\arabic{figure}}}
\section{Seasonal components extracted using Singular Spectrum Analysis\label{App:B}}	
\begin{figure}[H]
	\centering
	\includegraphics[width=1\textwidth]{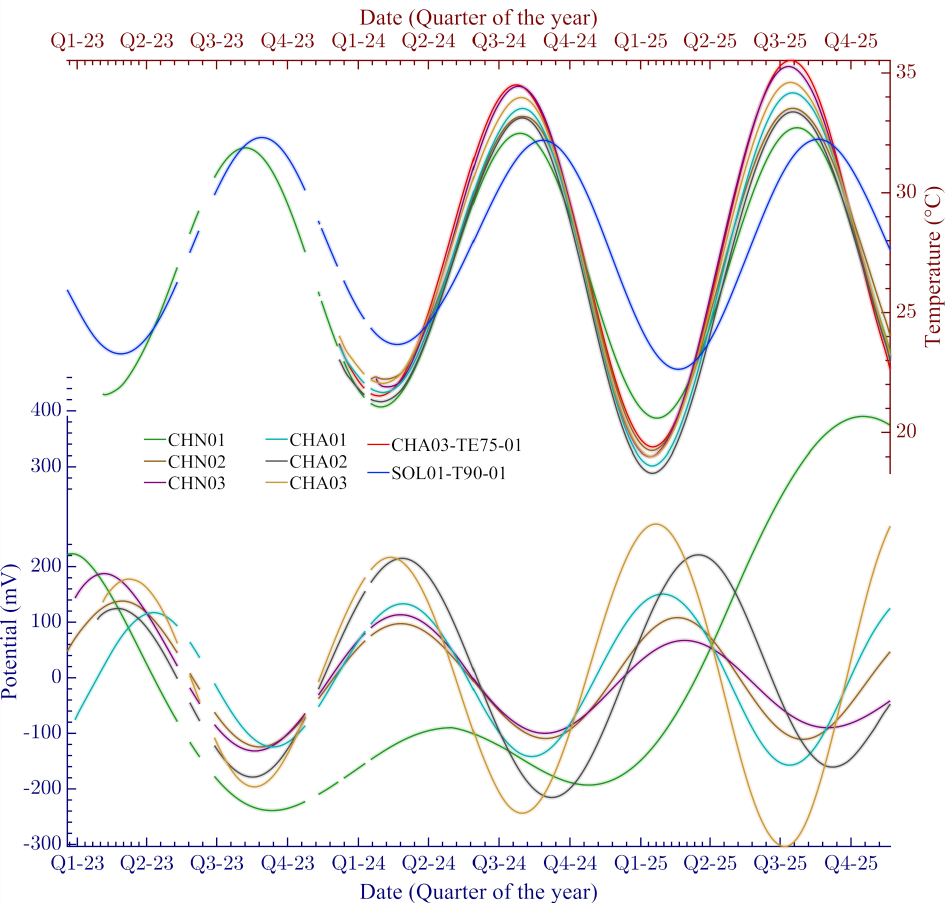}  
	\caption{Seasonal components extracted using Singular Spectrum Analysis (SSA).
		SSA-derived annual-scale components of trunk temperature (upper panel) and spontaneous electrical potential (lower panel) for the six monitored trees (three oaks and three hornbeams) over the 2023-2025 period. The extracted components isolate the dominant seasonal variability while preserving inter-individual differences. Short discontinuities correspond to acquisition interruptions.}
	\label{Fig:S02}
\end{figure}

To isolate the dominant annual-scale variability, Singular Spectrum Analysis (SSA) was applied separately to each daily temperature and spontaneous-potential time series. In practical terms, SSA was used to separate the slowly varying seasonal cycle from shorter-term fluctuations without imposing a sinusoidal shape or assuming identical seasonal dynamics among trees.

Before decomposition, each time series was centred using the mean of the available observations. Missing values were not interpolated. Instead, the lagged covariance was estimated at each time lag using only pairs of valid observations. These covariance estimates were used to construct a symmetric Toeplitz covariance matrix, which was decomposed into eigenvalues and empirical orthogonal functions. Principal components were calculated by projecting the lagged data vectors onto the corresponding empirical orthogonal functions while ignoring missing entries. The reconstructed modes were obtained by diagonal averaging, and positions that were missing in the original records were retained as missing values after reconstruction.

The reconstructed series shown in Figure~\ref{Fig:S02} emphasize the dominant annual-scale oscillations while preserving differences in amplitude and phase among individuals. No interpolation, additional filtering, or detrending was applied after SSA reconstruction.

\newpage 
\setcounter{figure}{0}
\renewcommand{\thefigure}{C\padzeroes[2]{\arabic{figure}}}
\section{Seasonal Lissajous diagrams of SP-temperature coupling across all monitored trees\label{App:C}}	
\begin{figure}[H]
	\centering
	\includegraphics[width=1\textwidth]{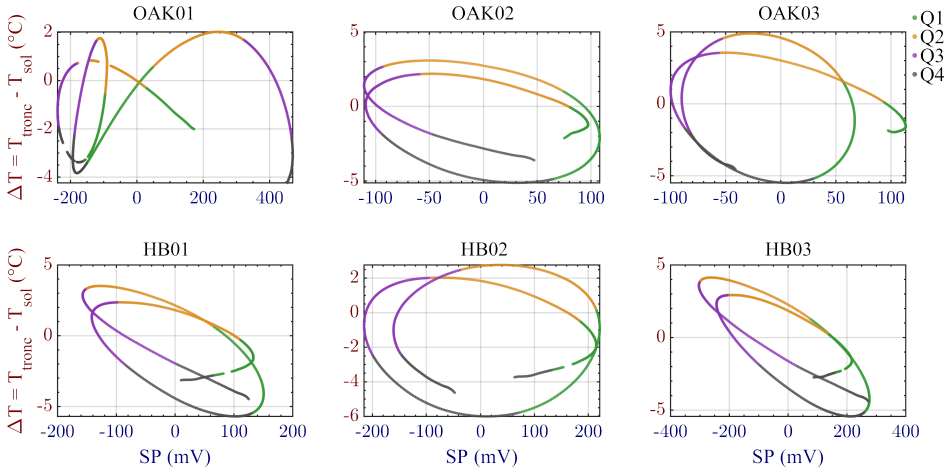}  
	\caption{Seasonal Lissajous diagrams for all monitored trees. Lissajous representations of the seasonal components of self-potential (SP) versus trunk–soil temperature difference ($\Delta T = T_trunk - T_soil$) for the six monitored trees (OAK01-03 and HB01-03). Seasonal signals were extracted using Singular Spectrum Analysis (SSA) (\cf Figure \ref{Fig:S02}, Appendix \ref{App:B}). Points are colored by calendar quarters (Q1-Q4). Across all individuals, structured elliptical or distorted-elliptical loops are observed, indicating a persistent seasonal phase lag between electrical and thermal dynamics.}
	\label{Fig:S03}
\end{figure}

Figure \ref{Fig:S03} presents the seasonal Lissajous diagrams for all monitored trees. In each case, the seasonal components extracted via Singular Spectrum Analysis (SSA) reveal coherent closed or quasi-closed loops in the SP-$\Delta T$ phase space. Although the amplitude and geometric distortion vary among individuals, a common feature emerges; the presence of a stable seasonal hysteresis between electrical potential and thermal forcing. The consistent orientation of the loops suggests a systematic phase lag, indicating that SP does not instantaneously follow thermal variations but responds with delayed dynamics. Inter-individual differences likely reflect variability in soil conditions, root architecture, hydraulic functioning, and local hydrothermal coupling. However, the persistence of structured phase trajectories across all trees supports the robustness of the observed thermo-electrical coupling at seasonal scale.

\newpage 
\setcounter{figure}{0}
\renewcommand{\thefigure}{D\padzeroes[2]{\arabic{figure}}}
\section{Seasonal instantaneous phase lag between SP and trunk-soil temperature difference\label{App:D}}	
\begin{figure}[H]
	\centering
	\includegraphics[width=1\textwidth]{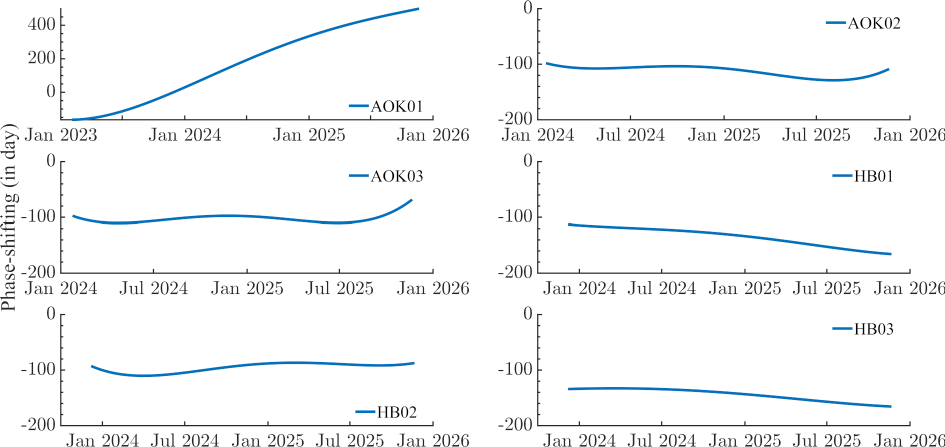}  
	\caption{Seasonal instantaneous phase lag between spontaneous electrical potential (SP) and trunk–soil temperature difference ($\Delta T$) for all trees, estimated using the Hilbert transform of SSA-smoothed signals. For five of the six trees (OAK02, OAK03, HB01, HB02, HB03), the phase lag remains consistently close to -100 days throughout the annual cycle, with limited intra-seasonal variability. OAK01 displays a distinct long-term drift and is not shown on the same temporal scale. The broadly similar offset observed in five individuals indicates a recurrent delayed seasonal relationship between the electrical and thermal signals, without identifying a unique underlying mechanism.}
	\label{Fig:S04}
\end{figure}

The instantaneous phase relationship between SP and trunk-soil temperature difference was estimated for each tree using the Hilbert transform applied to the SSA-reconstructed seasonal components. Five of the six trees exhibited a broadly similar seasonal offset, with phase lags remaining close to $-100$~days over much of the annual cycle (\cf Figure~\ref{Fig:S04}). The recurrence of this offset indicates that the electrical and thermal signals were governed by slowly varying seasonal processes rather than by synchronous short-term responses. The lag should not be interpreted as the travel time of water or heat through the soil--root--stem continuum, and its sign depends on the phase convention adopted. OAK01 showed a distinct long-term drift and a less stable phase relationship, highlighting inter-individual variability in the seasonal dynamics.

\end{document}